\documentclass{article}
\usepackage{spconf,amsmath,graphicx}
\usepackage{multirow}
\usepackage{amssymb}
\usepackage{pifont}
\usepackage{makecell}
\usepackage{arydshln}
\newcommand{\xmark}{\ding{55}}
\usepackage{tikz}
    \usetikzlibrary{calc}
\newlength{\Aheight}
\newcommand{\inline}[2]{%
    \begin{tikzpicture}[baseline = (word.base), txt/.style = {inner sep = 4pt, text height = \Aheight}, above/.style = {inner sep = 3pt, text depth = 5pt}]
    \node[txt] (word) {#1};
    \node[above] at (word.north) (label) {\footnotesize{#2}};
    \end{tikzpicture}%
    }

\title{END-TO-END ARCHITECTURES FOR ASR-FREE SPOKEN LANGUAGE UNDERSTANDING}
\name{Elisavet Palogiannidi, Ioannis Gkinis, George Mastrapas, Petr Mizera, Themos Stafylakis}
\address{Omilia - Conversational Intelligence, Athens, Greece \\
\{epalogiannidi, igkinis, gmastrapas, petr.mizera, tstafylakis\}@omilia.com}
\begin{document}
%\ninept
%
\maketitle
\begin{abstract}
Spoken Language Understanding (SLU) is the problem of extracting the meaning from speech utterances. It is typically addressed as a two-step problem, where an Automatic Speech Recognition (ASR) model is employed to convert speech into text, followed by a Natural Language Understanding (NLU) model to extract meaning from the decoded text. Recently, end-to-end approaches were emerged, aiming at unifying the ASR and NLU into a single SLU deep neural architecture, trained using combinations of ASR and NLU-level recognition units. In this paper, we explore a set of recurrent architectures for intent classification, tailored to the recently introduced Fluent Speech Commands (FSC) dataset, where intents are formed as combinations of three slots (action, object, and location). We show that by combining deep recurrent architectures with standard data augmentation, state-of-the-art results can be attained, without using ASR-level targets or pretrained ASR models. We also investigate its generalizability to new wordings, and we show that the model can perform reasonably well on wordings unseen during training. 
\end{abstract}
\section{Introduction}
\label{sec:intro}

Over the past decades voice interfaces have become an integral part of daily life communication for a wide range of domains like banking, entertainment or travelling making the development of Spoken Language Understanding (SLU) systems necessity.

Typical SLU systems employ an ASR module for decoding speech into text and NLU for estimating the meaning of the utterance, in the form of domain classification, intent classification, and slot filling \cite{tur2011spoken, bapna2017sequential}. The problem with these approaches is that the ASR errors are propagated to the next components of the SLU pipeline. A way to address this problem is by replacing the single decoding (1-best hypothesis) with word lattices or word confusion networks \cite{hakkani2006beyond}. Another approach is to utilize word embeddings that model acoustic relationships between words in addition to the semantic and syntactic relations \cite{shivakumar2019spoken}. 
An alternative approach is to employ end-to-end architectures capable of learning how to map sequences of acoustic features directly to SLU recognition units \cite{wang2006speech,qian2017exploring, chen2018spoken, haghani2018audio}. SLU units that are typically used are combinations of ASR-level units (e.g. phonemes, characters, word-pieces, words) with NLU-level units (e.g. intents, slots) \cite{serdyuk2018towards,tomashenko2019investigating}. Two-step training approaches have also been proposed, where the network is pretrained on large datasets using ASR-level recognition units, and it is subsequently finetuned on the target dataset using NLU-level recognition units \cite{chen2018spoken, lugosch2019speech}.

Intent classification is a crucial task for establishing a successful communication between the systems and the endpoint users \cite{firdaus2018deep}. 
A system that detects intents correctly is able to understand the purposes of the users and interact in the most appropriate way. Most systems handle intents as simple classes, but looking a step further, intents can be handled as structure or their names can be something more that labels, e.g., they can convey semantics. The NLU model of \cite{zhang2018joint} aims to encapsulate the hierarchical relationship among word, slot and intent, while the joint training on slot filling and intent classification has been proven to invigorate the NLU components \cite{zhang2018joint,bapna2017sequential,goo2018slot}.

In this work, we propose an end-to-end ASR-free architecture for intent classification, in the sense that it does not make any use of ASR-level recognition units (e.g. phonemes, characters, words) during training or evaluation. The architecture handles the intent as a structure of slots, by combining the predictions of each spot. We also conduct a series of experiments in order to investigate our architecture's degree of generalizability to unseen wordings.

\section{SLU Architecture}
\label{sec:models}
\subsection{Modeling assumptions}
The proposed SLU architecture is trained to predict intents directly from acoustic features. Rather than considering intents as classes, it handles them as tuples of slots, each having an associated SoftMax layer. 
The intent tuple depends on the dataset, and in the case of  \cite{lugosch2019speech} a three-slot tuple is defined by \textit{action}, \textit{object} and \textit{location}.
Following such an approach, the single-label classification task (intent) with a large number of classes is translated to a multi-class classification task (slots) with reduced number of classes. An intent is predicted correctly, if all three slots have been predicted correctly. 
One way of defining the intent probability is as follows:
\begin{equation}\label{eq:1}
    p(A,O,L|D) = p(A|D)\cdot p(O|D)\cdot p(L|D),
\end{equation}

\noindent where $A$, $O$, $L$ stand for action, object and location slots respectively, and $D$ for the sequence of acoustic features of the utterance. Note that Eq. (\ref{eq:1}) assumes conditional independence between slots given $D$. An alternative way of defining the intent probability -and more consistent with probability theory- is by assuming conditional dependence between slots as shown below:
\begin{equation}\label{eq:cond}
    p(A,O,L|D) = p(A|D)\cdot p(O|A,D)\cdot p(L|A,O,D).
\end{equation}
Note that any ordering of $\{A,O,L\}$ is valid. The conditional classifier consists of one independent and two dependent slots. The independent slot, similarly with a conditional independent approach, is predicted given $D$. On the other hand, the dependent slots are predicted given a composite representation that contains not only $D$, but information from the other slot(s) as well. 

\subsection{Training and inference}
The overall loss function is the summation of the three slot-specific cross entropy losses. Training the unconditional model is straightforward, since the model assumes conditional independence. During inference, constraints can be posed to ignore tuples that do not form a valid intent.
On the other hand, training the conditional model can be performed using teacher-forcing or scheduled sampling, that are commonly employed for training sequence-to-sequence models \cite{bengio2015scheduled}. During inference, one may consider Beam Search in order to maximize the $p(A,O,L|D)$, although in our experiments with FSC dataset we observed no significant gains compared to greedy optimization. 

\subsection{Architectural details}

Fig. \ref{fig:res} illustrates an overview of the end-to-end Recurrent Neural Network (RNN) based proposed architecture. It is comprised of three parts: (1) RNN-stack, (2) Representation-Layer and (3) Conditional classifier. 
% Mel Frequency Cepstral Coefficients (MFCCs) are used as acoustic features.
\vspace{-1em}
\subsubsection{RNN stack}
The RNN-stack may consists of $N$ Bidirectional or Unidirectional RNN layers.
Stacking multiple RNN layers makes the architecture deeper and enables the model to learn more complex input patterns.
Connections between stack's layers may be a) sequential i.e., the output of the layer $L_n$ is the input to the layer $L_{n+1}$ etc., or b) residual i.e., the input to the layer $L_{n+1}$ is the summation of the output of $L_n$ and $L_{n-1}$ layers.  The output of the stack is fed to the Representation-Layer.
\vspace{-1em}
\subsubsection{Representation layer}
%The Representation-Layer has been designed in order to have the flexibility to calibrate the representation that receives from the RNN-stack, and to formulate accordingly the input to the SoftMax Layer. It can be considered as an RNN-stack extension, which allows the insertion of a different RNN layer or a totally different RNN structure. 

The representation layer is responsible for squeezing the temporal information to enable sequence classification. It contains another RNN layer followed by average pooling.
Apart from a single RNN, it may contain a triple RNN structure. The latter consists of one RNN for each slot followed by average pooling. It aims at extracting three slot-specific representations, as opposed to a single representation for all three SoftMax classifiers.

% In this case, Representation-Layer diffuses the RNN-stack representation to one LSTM per slot that is connected with the corresponding classifier in the SoftMax Layer. We denote this structure as triple-LSTM.
\vspace{-0.5em}
\subsubsection{Classifier}

In the unconditional model, the three slots are predicted given the representation layer's output representation $r$. In the conditional model, assuming teacher-forcing training, the $object$ slot $s_o$ is predicted given $r$ and the $action$ ground-truth slot $s_a$. The input to the $location$ slot's classifier is formulated in a similar manner, given $r$ and ground-truth $s_a$, $s_o$. Conditioning e.g. on $s_a$ is implemented by concatenating $r$ with the column of the linear layer corresponding to the ground-truth label of $action$. When scheduled sampling is employed, conditioning is implemented by alternating between ground-truth and estimated slots \cite{bengio2015scheduled}. When triple-RNN is employed, the representation layer extracts three slot-specific representations, i.e. $r_a$, $r_o$ and $r_l$. 
Finally, the predicted intent is formulated by combining the predictions of action, object and location.

\begin{figure}[htb]

\begin{minipage}[b]{1.0\linewidth}
  \centering
  \centerline{\includegraphics[scale=0.64]{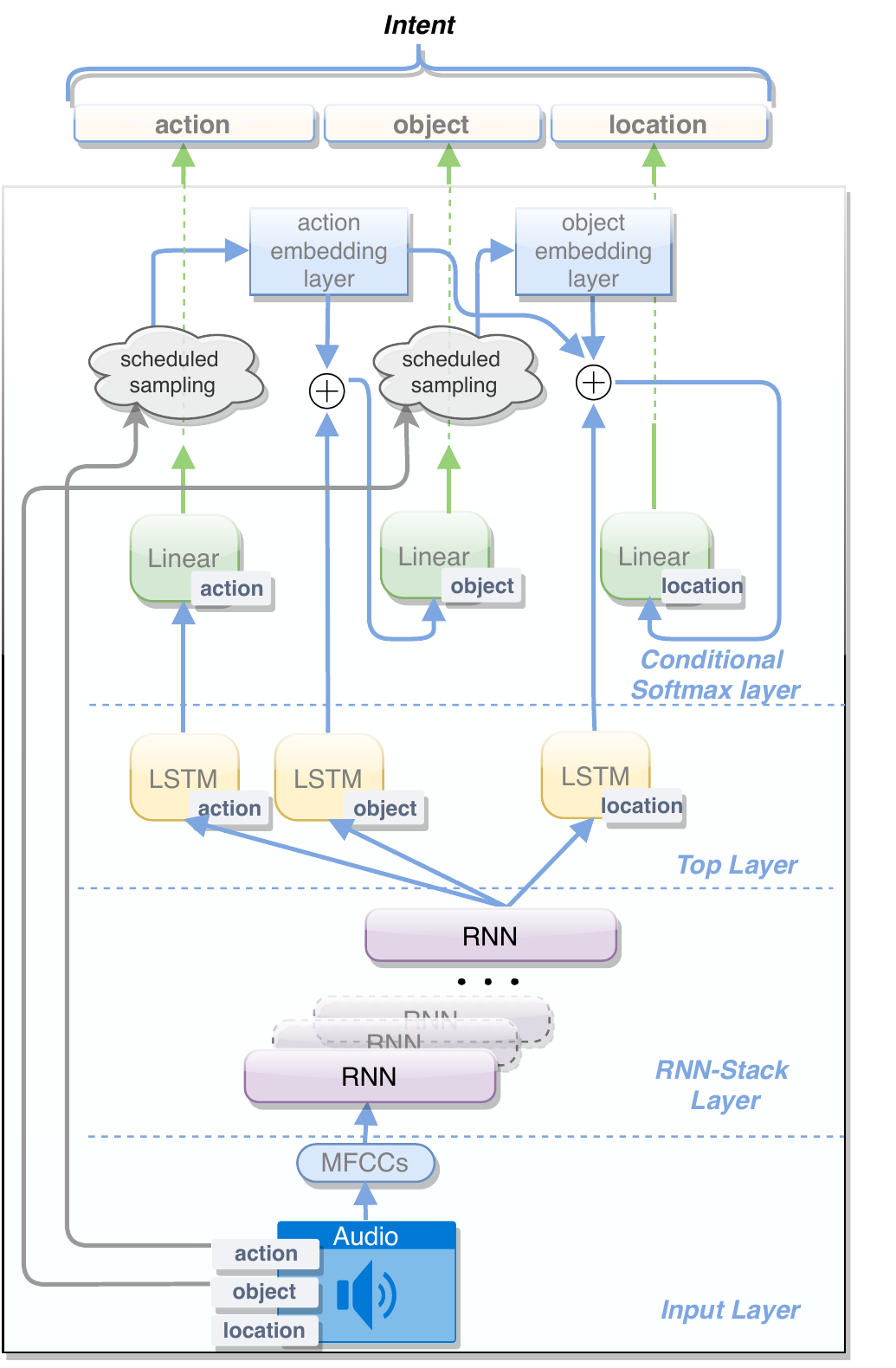}}
\end{minipage}
\caption{End-to-end SLU architecture}
\label{fig:res}
\end{figure}

\section{Experiments}
\label{sec:exp}
\subsection{Dataset and data augmentation}
The proposed models are trained from scratch and evaluated on the Fluent Speech Commands (FSC) dataset \cite{lugosch2019speech}. In order to avoid overfitting and to increase models' robustness, we train our architectures using five times the original train set, which we generate using Kaldi's data augmentation \cite{povey2011kaldi}. Data augmentation works as a regularization mechanism preventing the model from overfitting. The augmented training set is generated by applying four different data augmentation techniques: Reverberation, Music, Babble and Noise injection, where the noises are extracted from the MUSAN corpus \cite{snyder2015musan}. After applying the data augmentation, the number of training utterances increases from 23132 to 115660. In Table \ref{tab:dstats} the FSC dataset statistics are summarized.  
%For reverberation, three databases that consist of 325 real RIRs: the RWCP sound scene database \cite{kinoshita2013reverb}, the REVERB challenge database \cite{nakamura2000acoustical} and the Aachen impulse response database \cite{jeub2009binaural} were used. For the rest techniques the audio segments were diversified using speech, music and noise portion of the Musan corpus, respectively, which contains 109 hours of audio \cite{snyder2015musan}. 

Some examples of wordings and their intents are depicted in Table \ref{tab:examples}. If it is not obvious that a slot is associated to specific words, the slot value is set to $none$. Sometimes the slots can be inferred by the full context of the wording, e.g., the $action$ of  ``\emph{Too loud}" wording is $decrease$.
Some slots, like $heat$ can be expressed by semantically similar words like \emph{heat} and \emph{temperature}. In such cases, it is hard to understand that both words refer to the same slot without pretrained word embeddings. The same holds for phonetically similar words, i.e., ``\emph{on, off}". 

\begin{table}
  \centering
\begin{tabular}{|l||l|}
    \hline
    {\bf Train:} (Utterances, Speakers) & (115660, 77) \\\hline
    {\bf Validation:} (Utterances, Speakers) & (3118, 10) \\\hline
    {\bf Test:} (Utterances, Speakers) & (3793, 10) \\\Xhline{2\arrayrulewidth}
    {\bf Unique Intents} & 31\\\hline
    {\bf  Unique:} (Actions, Objects, Locations) & (6, 14, 4)\\\hline
  \end{tabular}
  \caption{FSC dataset statistics.}
  \label{tab:dstats}
\end{table}

\begin{table}
  \centering
\begin{tabular}{l}\\\hline
 \inline{Bathroom}{$location$: washroom} \inline{heat}{$object$: heat} \inline{up}{$action$: increase} \\\hline
   \inline{Bathroom}{$location$: washroom} \inline{heat}{$object$: heat} \inline{down}{$action$: decrease}\\\hline
  \inline{Decrease}{$action$: decrease} the \inline{heating}{$object$: heat} in the \inline{kitchen}{$location$: kitchen}\\\hline
   \inline{Increase}{$action$: increase} the \inline{temperature}{$object$: heat} in the \inline{bedroom}{$location$: bedroom}\\\hline
   \inline{Bring}{$action$: bring} me the \inline{newspaper}{$object$: newspaper} \inline{}{$location$: none}\\\hline
%   \inline{Bring}{$action$: bring} my \inline{shoes}{$object$: shoes}\inline{}{$location$: none}\\\hline
   \inline{Lights}{$object$: lights} \inline{on}{$action$: activate} \inline{}{$location$: none}\\\hline
    Turn the \inline{lamp}{$object$: lamp} \inline{off}{$action$: decrease} \inline{}{$location$: none}\\\hline
  \inline{Switch language}{$action$: change language} \inline{}{$object$: none} \inline{}{$location$: none}\\\hline
  \inline{Set my device}{$action$: change language} to \inline{Chinese}{$object$: Chinese}\inline{}{$location$: none}\\\hline
  \inline{Too loud}{$action$: decrease}\inline{}{$object$: volume} \inline{}{$location$: none}\\\hline
 
  \end{tabular}
  \caption{Examples of wordings and their intents. }
  \label{tab:examples}
\end{table}

\vspace{-1em}
\subsection{Architecture configuration}

The architecture shown in Fig. \ref{fig:res} can generate multiple models by configuring each layer. In the simplest case, RNN-stack consists of one layer, the representation layer is skipped, and the classifier consists of three SoftMax layers. In a more complex model the RNN-stack can consist of three layers, representation layer contains a triple-LSTM and the classifier is conditional. 

The models are trained for about 15 epochs on the FSC augmented training dataset and evaluated on the validation set in order to select the best model. The final evaluation, that is reported on the results occurs on the unseen data (test). We experimented with both LSTM-stack and GRU-stack, and with both sequential or residual connections. The representation layer can be a single LSTM, a single GRU or a triple LSTM. Bidirectional RNN with hidden size 512 were utilized. The initial learning rate is set to 0.001 and the scheduler halves it for the first time at the seventh epoch and then at every two epochs. A dropout layer is applied at the output of the representation layer and additional dropout layers are applied before the input to the classifiers. Moreover, batch normalization and average pooling on the representation layer's output were also utilized. The conditional models are trained with scheduled sampling \cite{bengio2015scheduled}. The probability of selecting ground-truth labels starts from 1.0 and decreases with the number of epoch using a sigmoid-shaped function until it reaches 0.5.

We use 40-dimensional MFCCs as acoustic features, extracted every 10ms. Cepstral mean normalization is applied, which statistics estimated from the whole utterance.
\vspace{-1em}
\subsection{Generalizing to new wordings}

Aiming to investigate how our architecture responds to test data with unseen wordings we conduct a series of experiments where we remove from the training data a set of wordings and then we evaluate the model only to wordings unseen during training. We create three training set versions by randomly choosing and removing from the training dataset 20 or 50 out of the 248 unique wordings. We also set up an extreme case, in which we keep in the training set only the most frequent wording. In every case, we make sure that the unique intents set do not change. In the Table \ref{tab:dsize} we list the new sizes of the  train and the test datasets. 

% \begin{table}
%   \centering
% \begin{tabular}{|c||c|c|c|}
%   \Xhline{2\arrayrulewidth}
%     \multirow{2}{*}{ \makecell{\bf Unique wordings \\ \bf in  train set}}  & \multirow{2}{*}{\bf Train Size} & \multicolumn{2}{|c|}{\bf Test utterances} \\\cline{3-4}
%      &  & \bf Unseen & \bf Seen \\ \Xhline{2\arrayrulewidth}
%     248   & 115660 & -&3793 \\%\hline
%     \Xhline{2\arrayrulewidth}
%     228   & 106360   & 306 & 3487 \\%\hline
%     \hline
%     198   & 92885  & 745 & 3048 \\%\hline
%     \hline
%     31    & 17650  & 3228 & 565\\\hline
%   \end{tabular}
%   \caption{Train and test dataset sizes used for generalization. The first line corresponds to the original dataset.}
%   \label{tab:dsize}
% \end{table}

\begin{table}
  \centering
\begin{tabular}{|c||c|c|c|c|c|}
  \Xhline{2\arrayrulewidth}
    \multirow{2}{*}{ \makecell{\bf Unique \\ \bf wordings}}  & \multirow{2}{*}{ \makecell{\bf Train \\ \bf size}} & \multicolumn{2}{|c|}{\bf Test utterances} & \multicolumn{2}{|c|}{\bf Error (\%)} \\\cline{3-6}
     &  & \bf Unseen & \bf Seen & \bf Unseen & \bf Seen \\ \Xhline{2\arrayrulewidth}
    248   & 23132 & -&3793 &-  & 1.49\\%\hline
    \Xhline{2\arrayrulewidth}
    228   & 21272   & 306 & 3487 & 15.25 & 1.81 \\%\hline
    \hline
    198   & 18577  & 745 & 3048 & 22.45 & 1.59\\%\hline
    \hline
    31    & 3530  & 3228 & 565 & 57.08 & 2.13\\\hline
  \end{tabular}
  \caption{Train and test dataset sizes used for generalization. The first line corresponds to the original dataset.}
  \label{tab:dsize}
\end{table}

\noindent Each reduced train dataset is accompanied with two test sets, one that contains only the unseen wordings and its complementary set, containing those seen during training.

\vspace{-1em}
\subsection{Experimental Results}

For each architecture, the best epoch is selected based on the performance on the validation set and the intent classification error achieved on the test set is reported in the table \ref{tab:res}. The results have been derived by training on the augmented dataset.

\begin{table}
  \centering
\begin{tabular}{|c|c|c||c|}
  \Xhline{2\arrayrulewidth}
    \bf RNN-stack & {\bf Repr. Layer} & {\bf Conditional} & {\bf Error (\%)}\\\Xhline{2\arrayrulewidth}
    1-LSTM   & \xmark & \xmark & 9.38 \\\hline
    2-LSTM   & triple-LSTM & \xmark & 2.48 \\\hline
    % 2-LSTM   & triple-LSTM & \xmark & 2.15 \\\hline
    \multirow{5}{*}{3-LSTM}   &\multirow{2}{*}{LSTM} & \xmark & 1.34 \\\cline{3-4}
                              & & \checkmark & 1.15 \\\cline{2-4}
                              & \multirow{2}{*}{triple-LSTM} & \xmark & 1.76 \\\cline{3-4}
                              & & \checkmark & 1.49 \\\cline{2-4}
                              & \multirow{2}{*}{GRU} & \xmark & 1.48 \\\cline{3-4}
                              & & \checkmark & 1.28 \\\hline
    3-GRU    & triple-LSTM & \xmark & 2.54 \\\hline\hline
    \multicolumn{3}{|c||}{State-of-the-art without pretraining \cite{lugosch2019speech}} & 3.40\\\hline
     \multicolumn{3}{|c||}{State-of-the-art with pretraining \cite{lugosch2019speech}} & 1.20\\\hline
    % 4-LSTM   & triple-LSTM & \xmark & 1.24 \\\hline

  \end{tabular}
  \caption{Intent classification error per model.}
  \label{tab:res}
\end{table}

Examining the results in the Table \ref{tab:res} we observe that all models employing a 3-LSTM stack attain performance comparable to the state-of-the-art. As presented in \cite{lugosch2019speech}, the state-of-the-art without for the FSC database is $3.40\%$ and with pretraining (on LibriSpeech) is $1.20\%$. The best performing model has triple-LSTM in the representation layer and a conditional classifier. Apart from the results reported here, other configurations such as 4-LSTM stack, residual RNN-stack connections, unidirectional RNNs, larger and smaller RNN sizes than 512 were also investigated, without however yielding results of any statistically significant differences. As regards the conditional model, the best conditioning order of slots found to be $action$-$object$-$location$, but with minor differences compared to $object$-$action$-$location$. We also investigated the impact of the data augmentation by training models both on clean and augmented dataset. Note that the best performing model without data augmentation attained classification error equal to 4.64\%. 
\vspace{-1em}
\subsection{Comparison with ASR-driven SLU system}

The performance of ASR-driven SLU systems was also investigated in \cite{SMansalis2019} and the results are summarized in table \ref{tab:asr_res}. The ASR model is a state-of-the-art Kaldi recipe (8-layer TDNNs, LF-MMI training \cite{povey2011kaldi,povey2016purely}). The architecture is trained on 1832 hours of publicly available datasets (Librispeech, WSJ, a.o.), while FSC is not included in the training set. The model attains  9.44\% WER on the FSC test set. The ASR-driven SLU models are trained either on the 1-best or on the N-best hypotheses (where N=10), employing transformer-based architectures such as BERT, RoBERTa and DistilBERT \cite{devlin2018bert,liu2019roberta,sanh2019distilbert}. N-best trained models outperform the corresponding 1-best and achieve state-of-the-art performance, slightly inferior to our end-to-end conditional LSTM model.
\subsection{Results on unseen wordings and discussion}
In Table \ref{tab:dsize} we summarize the results on wordings unseen during training and the corresponding statistics on training and test sets. The error rates indicate that the proposed architecture is to some extent capable of generalizing to new wordings and attaining decent performance, without further tuning. As expected, error rates on the unseen test increase with the number of wordings removed from the training set. We consider this as an inherent limitation of end-to-end ASR-free SLU approaches. State-of-the-art NLU approaches, based either on word embeddings or BERT-like architectures are trained on massive textual corpora and hence they are capable of generalizing easily to new wordings \cite{devlin2018bert,pennington2014glove}. Contrarily, our end-to-end SLU method is trained from scratch on FSC, and has no obvious mechanism for incorporating e.g. word embeddings. As a result, it can rely only on training sets containing several wordings per slot or intent for attaining state-of-the-art generalizability.

% All models that used the N-best lists for training outperform the ones used 1-best hypothesis, with DistilBERT being the dominant one. Although, the results of this work are comparable to the the state-of-the-art, our triple-LSTM with a conditional classifier configuration outperforms their DistilBERT. 

%However the same does not happen on the known test sets, since the error has been increased by less than 1\% on a model that has been trained to a dataset that is approximately six times smaller than the full augmented dataset. Evaluating only to the unseen wordings introduces extra difficulty, however, results indicate potential for further investigation.

% \begin{table}
%   \centering
% \begin{tabular}{|c||c|c|}
%   \Xhline{2\arrayrulewidth}
  
%   \multirow{2}{*}{ \makecell{\bf Unique wordings \\ \bf in  train set}}  & \multicolumn{2}{|c|}{\bf Error (\%)} \\\cline{2-3}
%      &  \bf Unseen & \bf Seen \\\Xhline{2\arrayrulewidth}
%     248  &-  & 1.38 \\\hline
%     228  & 15.25 & 1.81 \\\hline
%     198  & 22.45 & 1.59\\\hline
%     31   & 57.08 & 2.13\\\hline
%   \end{tabular}
%   \caption{Intent classification error on seen and unseen wordings.}
%   \label{tab:gen}
% \end{table}

\begin{table}
  \centering
\begin{tabular}{|l||c||c|}
  \Xhline{2\arrayrulewidth}
    \bf Model & {\bf 1-best (\%)} & {\bf N-best (\%)} \\\Xhline{2\arrayrulewidth}
    BERT (base) & 3.69 & 3.62 \\\hline
    RoBERTa (base) & 3.87 & 3.34 \\\hline
    DistilBERT & 1.64 & 1.44 \\\hline
  \end{tabular}
  \caption{Intent classification error per ASR hypothesis \cite{SMansalis2019}.}
  \label{tab:asr_res}
\end{table}

\section{Conclusions and Future work}
\label{sec:concl}

In this work we investigated end-to-end RNN-based SLU architectures for intent classification, without the need of any ASR supervision. We explored several architectural variations and experimented with the use of conditional prediction of the slots composing the intents. Additionally, we demonstrated that data augmentation is compulsory for attaining state-of-the-art performance and training deep architectures. We also investigated the generalizability of the proposed model to wordings unseen during training, and found that it can attain fairly good results on such wordings. 

%The models achieve high performance even if they are trained on a significantly smaller training dataset, and given that training dataset is large enough, the performance can relatively tolerant in unseen data as well.
%Our architecture, compared with the corresponding model without pre-training presented in \cite{lugosch2019speech} reduces the error by approximately 2\%.

In the future we plan to train and evaluate the architecture on other SLU datasets, enhance the architecture with attention mechanism, and replace the recurrent layers with convolutional or attentive layers.

\bibliographystyle{IEEEbib}
\bibliography{refs}

\end{document}